\documentclass[%
 aip,
 amsmath,amssymb,
 reprint,%
]{revtex4-2}

\usepackage[pdftex]{graphicx}
\usepackage{amsthm}
\usepackage{dcolumn}
\usepackage{bm}
\newtheorem{theorem}{Theorem}
\newtheorem{corollary}[theorem]{Corollary}
\newtheorem{proposition}[theorem]{Proposition}
\usepackage[utf8]{inputenc}
\usepackage[T1]{fontenc}
\usepackage{mathptmx}

\usepackage{pgfplots,subfig}
\pgfplotsset{ 
  compat=newest, 
  colormap={CM}{rgb=(.6,0,0) rgb=(1,0,0) color=(white) color=(blue) rgb=(0,0,0.6)},
every tick label/.append style={font=\small},
every axis label/.append style={font=\small}}
\usepgfplotslibrary{fillbetween}
\usetikzlibrary{shapes.multipart}

\usepackage{epstopdf}
\usepackage{epsfig}
\definecolor{mygreen}{RGB}{72,160,66}
\definecolor{myblue}{RGB}{20,155,204}
\definecolor{myyellow}{RGB}{190,170,0}
\definecolor{myviolet}{RGB}{195,8,255}
\definecolor{myred}{RGB}{204,0,0}
\definecolor{myblue}{RGB}{0,66,255}

\begin{document}

\preprint{AIP/123-QED}

\title[A two-layer model for coevolving opinion dynamics and collective decision-making in complex social systems]{A two-layer model for coevolving opinion dynamics and collective decision-making in complex social systems}

\author{Lorenzo Zino} \email{lorenzo.zino@rug.nl}
\affiliation{ 
Faculty of Science and Engineering, University of Groningen, 9747 AG Groningen, the Netherlands
}%
\author{Mengbin Ye}\email{mengbin.ye@curtin.edu.au}
\affiliation{ 
Optus--Curtin Centre of Excellence in Artificial Intelligence, Curtin University, Perth 6102, WA, Australia}%
\affiliation{ 
Faculty of Science and Engineering, University of Groningen, 9747 AG Groningen, the Netherlands
}%
\author{Ming Cao}\email{m.cao@rug.nl}
\affiliation{ 
Faculty of Science and Engineering, University of Groningen, 9747 AG Groningen, the Netherlands
}%

\date{\today}

\begin{abstract}
Motivated by the literature on opinion dynamics and evolutionary game theory, we propose a novel mathematical framework to model the intertwined coevolution of opinions and decision-making in a complex social system. In the proposed framework, the members of a social community update their opinions and revise their actions as they learn of others' opinions shared on a communication channel, and observe of others' actions through an influence channel; these interactions determine a two-layer network structure.  We offer an application of the proposed framework by tailoring it to study the adoption of a novel social norm, demonstrating that the model is able to capture the emergence of several real-world collective phenomena such as paradigm shifts and unpopular norms. Through the establishment of analytical conditions and Monte Carlo numerical simulations, we shed light on the role of the coupling between opinion dynamics and decision-making, and of the network structure, in shaping the emergence of complex collective behavior in social systems.
\end{abstract}

\maketitle

\begin{quotation}
Mathematical models have emerged as powerful tools to describe and study the behavior of complex social systems. Here, we focus on the emergent behavior of a social community whose members dynamically revise their opinion and take collective decisions.  Motivated by the empirical evidence of an interdepencency between these two social dynamical processes, and the lack of mathematical tools to effectively describe it, we establish a modeling framework for the interdependent coevolution of opinions and decisions, extending and unifying the separate literature bodies on dynamic opinion formation and collective decision-making. We specialize the model to offer a realistic application of the proposed framework in which we study the introduction of an advantageous innovation in a social community, and focus on the factors of coupling strength between the opinion dynamics and decision-making, and the network structure. Depending on how such factors combine, a range of different complex real-world phenomena can be captured in our framework, enabling us to elucidate whether the society will see a paradigm shift in which individuals overwhelmingly adopt the innovation, the emergence of an unpopular norm where individuals fail to adopt the innovation despite opinions being overwhelmingly in favor of it, or a community that persistently supports the status quo over the innovation.
\end{quotation}

\section{\label{sec:introduction}Introduction}

The use of mathematically- and physically-principled models to represent and study social systems has become increasingly popular in the last decades~\cite{castellano2009social,Shao2009percolation,ratkiewicz2010popularity,baumann2020echo}. Researchers from a wide range of communities, including physics, applied mathematics, systems and control engineering, computational sociology, and computer science have devoted their efforts to capture the complexity of collective behavior within mathematical models that allows one to accurately predict the evolution of a social system, shedding light on the role of the individual-level dynamics on the emergence of complex collective behavior at the population level.

Since the 1950s, mathematical models have been widely adopted in social sciences to capture the complex phenomena that may emerge when members of a community interact and share their opinions. Among the literature, we mention the seminal works by French, DeGroot, Friedkin and Johnsen, which paved the way for the development of the mathematical theory of opinion dynamics and social influence ~\cite{french1956_socialpower,degroot1974OpinionDynamics,friedkin1990_FJsocialmodel}. Recently, these classical works have been extended to incorporate features of complex networks, such as antagonistic interactions and the emergence of disagreement~\cite{altafini2013antagonistic_interactions,acemoglu2013disagreement}, bounded confidence~\cite{hegselmann2002opinion,lorenz2007opinion}, the external influence of media~\cite{quattrociocchi2014}, the heterogeneous and time-varying nature of human interaction patterns~\cite{zino2019opinion,hasanyan2020leader}, and the coevolution of opinions and network structure~\cite{Holme2006coevolution}.

Collective decision-making is another real-world phenomenon that has been extensively studied by means of mathematical models. Typically such models describe how an individual's decision between a set of possible actions evolves as he or she takes into account the decisions of other individuals that he or she interacts with on a social network. Since its formalization in the 1970s, evolutionary game theory has emerged as a powerful paradigm and sound modeling framework for collective decision-making in social communities~\cite{smith1982book,young2011dynamics,montanari2010spread_innovation,ramazi2016networks}.

The social-psychological literature provides clear evidence that the two processes of opinion dynamics and collective decision-making are deeply intertwined and readers may intuitively appreciate such a coupling. On the one hand, it is indisputable that an individual's opinion has a key role in his or her decision-making process. On the other hand, many social-psychological theories support that the converse is also often observed; the actions that one individual observes from the others can shape  his or her opinion formation process. Existing literature reporting this include the social intuitionist model~\cite{haidt2001emotional}, norm interiorization processes~\cite{gavrilets2017interiorization}, as well as experimental evidence~\cite{lindstrom2018role}. Surprisingly, few efforts have been made toward generating a rigorous modeling framework for complex social systems with such a coupled coevolution of opinion and decision-making dynamics. We mention some works in which actions are modeled as quantized outputs
of the individuals' own opinion, evolving independently of others' actions~\cite{gargiulo2012OpDyn_Game,martins2008continuous}. Other efforts assume that each individual has a private opinion that is fixed and influences his or her decision-making process~\cite{centola2005selfenforce_norms}, that may vary according to external factors with a decision-making process that coevolves with the network structure~\cite{Schleussner2016coevolution}, or that coevolves along with an expressed opinion, but in the absence of a decision-making process~\cite{duggins2017_psych_opdyn,ye2019_EPO}.

Motivated by these preliminary works, the first key contribution of this paper is the development of a general modeling framework for the coevolution of opinion dynamics and collective decision-making in complex social systems. In the proposed model, the opinion dynamics of an individual evolves not only as a consequence of opinion sharing with other individuals, but also due to the influence from observing the actions of other individuals. The individual's decision-making process is governed by a coordination game to select between two alternative actions~\cite{young2011dynamics}, which is a classical framework to model the social tendency to conform with the actions of others, but is now additionally shaped by the individual's own opinion. In general, the opinion sharing process and the social influence from observed actions can occur between different pairs of individuals, and follow diverse interaction patterns. For this reason, we define our coevolutionary model on a two-layer social network, where a \emph{communication layer} is used to represent how individuals share their opinion, and an \emph{influence layer} captures the social influence due to observing the actions of others. Similar two-layer techniques have been used to represent epidemic processes and the simultaneous diffusion of awareness on the disease~\cite{granell2013multilayer,wang2019coevolution}, or to model complex synchronization dynamics~\cite{gambuzza2015multiplex,nicosia2017multiplex}.

In the last few years, several works have examined the key role played by the topology of a complex network in shaping the evolution of dynamical processes occurring on its fabric. Paradigmatic examples can be found in different fields, ranging from agreement dynamics and emergence of social power~\cite{barrat2007agreement,jalili2013opinion,liu2018leaders} to epidemic outbreaks in human populations~\cite{granell2013multilayer,pastorsatorras2015epidemics} and synchronization of power grids~\cite{rohden2014grids}. Besides deepening our understanding of the mechanisms that governs complex phenomena on networks, these results been used to inform the development of methodologies to control their evolution, such as in optimal vaccine allocation problems~\cite{nowzari2017}, or the implementation of pinning control to synchronize coupled oscillators~\cite{wang2002pinning,delellis2010pinning}.

In our second key contribution, we use the proposed modeling framework to study the effect of network topology on the formation of social norms, in particular focusing on the emergence of a paradigm shift (in which an innovation replaces the status quo norm), and on the phenomenon of unpopular norms~\cite{centola2005selfenforce_norms,willer2009false_enforcement,Smerdon2019}, in which a social community exhibits a collective behavior that is disapproved by most of the members of the community. A classical example is on alcohol abuse by college undergraduates in Princeton university campus at the beginning of the 1990s; it was found that even though most of the students were privately uncomfortable with the alcohol practices on campus but publicly continued to partake in heavy drinking~\cite{Prentice93pluralisticignorance}. We model the formation of social norms by studying the introduction of a social innovation in a community, represented as one of the two actions, and supported by a stubborn innovator individual~\cite{montanari2010spread_innovation}.

A theoretical result is derived for a necessary condition to observe the diffusion of the innovation when the decision-making process of all individuals is fully rational, dependent on the structure of the influence layer and on the role of an individual's opinion in his or her decision-making process. Then, we put forward an extensive simulation to study the case of bounded rational individuals. We find that the diffusion of the innovation is strongly influenced by the network structure and the coupling strength between the two coevolutionary dynamics. Phase transitions are identified between three different regions of the parameter space in which we observe i) a paradigm shift, ii) the emergence of an unpopular norm, and iii) the persistence of a popular but disadvantageous status quo, respectively. We demonstrate that the network structure plays a key role in determining in a nontrivial way the shape of these three regions and the sharpness of the phase transition between them. For instance, network topologies that seem to favor the occurrence of a paradigm shift when an individual's opinion is only slightly influenced by the actions of others, are instead strongly resistant to the introduction of the innovation when this influence increases in strength. 

The rest of the paper is organized as follows. In Section~\ref{sec:model}, we propose and discuss the coevolutionary modeling framework. In Section~\ref{sec:innovation}, we introduce our model for the adoption of innovation. Section~\ref{sec:results} is devoted to presenting our main findings. Section~\ref{sec:conclusion} presents discussion of our findings and outlines avenues for future research.

\section{\label{sec:model}Model}
In this section, we propose a novel modeling framework to capture the coevolution of the opinions and decisions of individuals interacting on a complex social network. After the model is formally introduced, we explain the intuition and motivation of the model by providing details on its components.

We consider a population of $n \geq 2$ individuals, indexed by the set $\mathcal{V} = \{1, \hdots, n\}$. Each individual $i \in \mathcal{V}$ is characterized by a two dimensional state variable $(x_i,y_i)$. The first component of the state variable represents a binary \emph{action} $x_i \in \{-1, +1\}$ made by the individual, while the second component $y_i \in [-1, 1]$, models his or her continuously distributed \emph{opinion}. The opinion measures the  individual's preference for an action, so that $y_i = -1$, $y_i=0$, and $y_i = 1$ represents that individual $i$ has maximal preference for action $-1$, is neutral, and has maximal preference for action $+1$, respectively. 

\begin{figure}
    \includegraphics{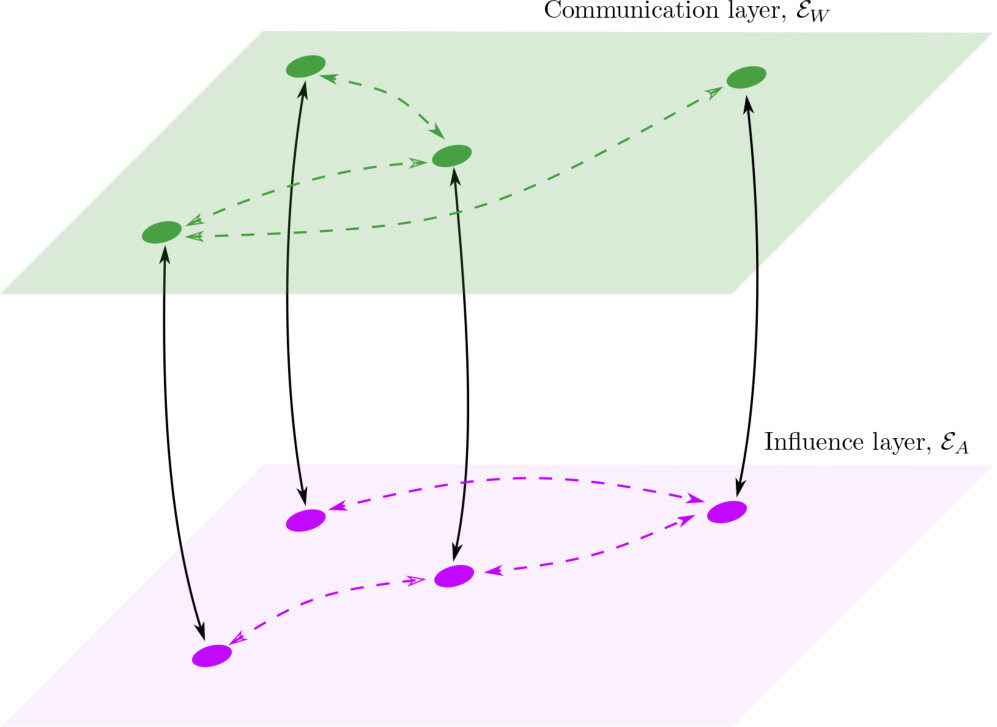}
    \caption{The coevolutionary dynamics occurs over a two-layer network. In the communication layer (in green), with edge set $\mathcal{E}_W$, individuals exchange opinions with each other. In the influence layer (in violet), with edge set $\mathcal{E}_A$, individuals are able to observe the actions of other individuals. The edge sets of the two layers are not necessarily the same. For example, an individual may choose to only share his or her opinion with a few close friends and family, but is able to observe and be influenced by the actions of many others in his or her community. Similarly, he or she may not be able to observe the action of individuals with whom he or she shares his or her opinion (e.g., due to long-distance interactions).}
    \label{fig:graph}
\end{figure}

The individuals update their actions and opinions after interacting with their peers on a two-layer network~\cite{kivela2014_multilayer}: the first layer models how individuals observe and are influenced by others' actions, while the second layer models how individuals communicate and exchange opinions with one another. We term these as the \emph{influence layer} and \emph{communication layer}, respectively. In general, the two layers are characterized by two different topologies, as illustrated in Fig.~\ref{fig:graph}. The influence layer is characterized by the undirected edge set $\mathcal{E}_A$, with an associated (unweighted) adjacency matrix $A\in\{0,1\}^{n\times n}$, having entries $a_{ij}$ defined as:
\begin{equation}
a_{ij}=\left\{\begin{array}{ll}1&\text{if }(i,j)\in \mathcal E_A,\\0&\text{if }(i,j)\notin \mathcal E_A.\end{array} \right.
\end{equation}
We assume that no self-loops are present, that is, all diagonal entries of $A$ are equal to $0$, and denote by 
\begin{equation}d_i = \vert \{(i, j) \in \mathcal{E}_A\} \vert\end{equation}
the degree of individual $i$ in the influence layer. 
The communication layer is characterized by the undirected edge set  $\mathcal{E}_W$ and a weighted adjacency matrix $W\in \mathbb R^{n\times n}$, with entries $w_{ij}\neq 0\iff (i,j)\in\mathcal E_W$. Self-loops are allowed in $\mathcal{E}_W$, and occurrence of a negative $w_{ij}$ would result in a signed network~\cite{altafini2013antagonistic_interactions}. Even though $\mathcal{E}_W$ is undirected, $W$ is not necessarily symmetrical, since $w_{ij}$ and $w_{ji}$ may be different. Although this work assumes that both layers are undirected, the proposed model easily admits a generalization to directed topologies on either layer, which may be investigated in future works.

The states of the individuals (i.e., opinions and decisions) evolve over discrete time-steps  $t = 0, 1, \hdots$. At each time $t$, a single individual $i\in\mathcal{V}$, selected uniformly at random and independently of the past history of the process\footnote{We observe that more realistic activation rules which account for temporal and inter-individual heterogeneity may be simply implemented by associating a (possibly inhomogeneous) Poisson clock to the activation of each individual.}, is activated and updates his or her opinion and action simultaneously, according to the following mechanisms.
\begin{description}
    \item[Opinion dynamics] the opinion of individual $i \in \mathcal{V}$ evolves as
\begin{equation}\label{eq:opinion}
    y_i(t+1) = (1-\mu_i)\sum_{j=1}^n w_{ij}y_j(t)+ \mu_i \frac{1}{d_i} \sum_{k=1}^n a_{ik} x_k(t),
\end{equation}
where the parameter $\mu_i \in [0, 1]$, called \emph{susceptibility}, measures the influence of his or her neighbors' actions $x_k(t)$ of the individual's opinion.
    \item [Decision making] the action of individual $i \in \mathcal{V}$ evolves according to a stochastic process. Specifically, the probability for individual $i$ to take action $x \in \{-1, +1\}$ at time $t+1$ is 
\begin{equation}\label{eq:loglinear}
    \mathbb{P}(x_i(t+1) = x) = \frac{e^{\beta_i\pi_i(x)}}{e^{\beta_i\pi_i(x)}+e^{\beta_i\pi_i(-x)}},
\end{equation}
where $\beta_i > 0$ measures the individual's rationality in the decision-making process, and $\pi_i(x) = \pi_i(x | y_i, x_{-i})$ is the payoff for individual $i$ to take action $x$, given his or her current opinion $y_i(t)$ and the actions of the others, $x_{-i}(t):= [x_{1}(t), \hdots, x_{i-1}(t), x_{i+1}(t), \hdots, x_n(t)]^\top\in\{-1,+1\}^{n-1}$. We define the following payoff function:
\begin{align}\label{eq:payoff}
    \pi_i(x\,|&\, y_i, x_{-i})  = \frac12\lambda_i x y_i \nonumber \\ 
    & + \frac{1-\lambda_i}{4d_i} \sum_{j=1}^n a_{ij}\begin{bmatrix} 1+x \\ 1-x\end{bmatrix}^\top\begin{bmatrix} 1+\alpha & 0 \\ 0 & 1\end{bmatrix}\begin{bmatrix} 1+x_j \\ 1-x_j\end{bmatrix}, 
\end{align}
where $\alpha \geq 0$ captures the evolutionary advantage of action $+1$ over action $-1$ and the parameter $\lambda_i \in [0, 1]$, called \emph{commitment}, measures the importance that individual $i$ gives to his or her own opinion in the decision-making process.
\end{description}

\begin{figure}
    \centering
    \includegraphics{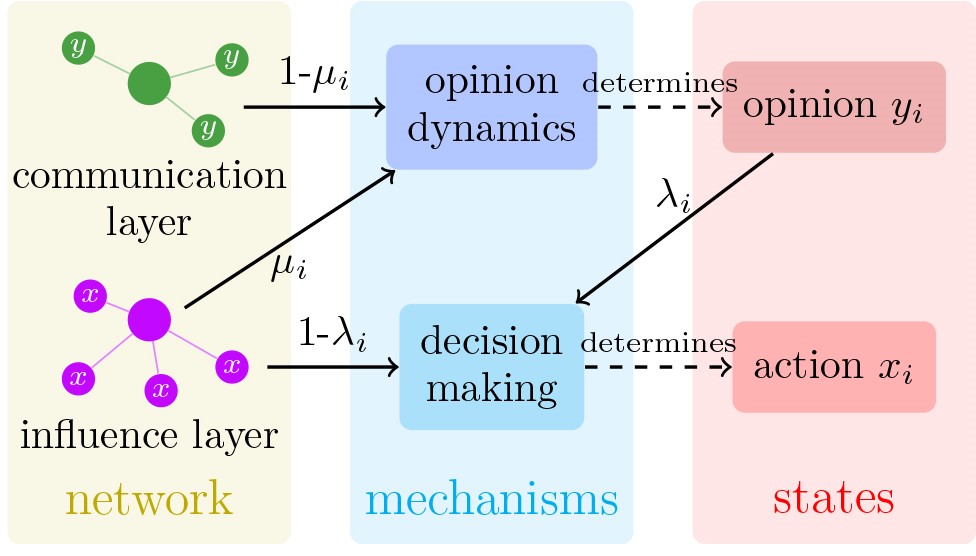}
    \caption{Schematic of the coevolutionary dynamics, and the interdependence between the two mechanisms.}
    \label{fig:schematic}
\end{figure}

The remainder of this section is devoted to a detailed discussion and motivation on the two intertwined components that compose the novel coevolutionary dynamics of opinions and decisions in our proposed model, which are illustrated in the schematic in Fig.~\ref{fig:schematic}.

\subsection{\label{ssec:opinion}Opinion Dynamics Component}

According to~\eqref{eq:opinion}, the opinion of individual $i$ at time $t+1$ is a convex combination of two summands: the first one accounts for the opinions of the individuals with whom individual $i$ interacts on the communication level; the second term captures the influence of the actions observed by individual $i$ on the influence level. Such a convex combination is regulated by the susceptibility $\mu_i\in[0,1]$, which measures the influence of the actions observed on the individual's opinion so that $\mu_i=0$ models the case the opinion evolves independently of the actions observed.

The products $(1-\mu_i)w_{ij}$ and $\mu_i a_{ij}$ are the weights that individual $i$ assigns to the opinion and action, respectively, of individual $j$. Since the two layers of the network may have different edge sets, it is in general possible that one of the two weights is nonzero and the other is zero. A standard assumption, which we shall adopt in the rest of this paper and often made in opinion dynamics models~\cite{proskurnikov2017tutorial}, is that $\sum_{j=1}  |w_{ij}| = 1$ for all $i\in\mathcal{V}$. This assumption guarantees that the opinions in the coevolutionary model are always well defined, as explicitly stated in the following result, whose proof is given in Appendix~\ref{app:opinion}.

\begin{proposition}\label{prop:well}
Let $W$ be such that  $\sum_{j=1}  |w_{ij}| = 1$, for all $i\in\mathcal{V}$, and let the initial opinions satisfy $y_i(0)\in[-1,1]$, for all $i\in \mathcal{V}$. Then, $y_i(t) \in [-1, 1]$, for all $i\in\mathcal{V}$ and $t\geq 0$.
\end{proposition}

If the weights of the communication layer are nonnegative, that is, $w_{ij} > 0$ for all $(i,j)\in\mathcal E_W$, then the updated opinion is a weighted average of i) the actions $x_k(t)$ of his or her neighbors on the influence layer, and ii) the opinions $y_j(t)$ of his or her neighbors on the communication layer. Beginning with the classical French--DeGroot model~\cite{french1956_socialpower,degroot1974OpinionDynamics}, weighted averaging is seen as a classical approach to modeling the way an individual processes, and is influenced by, external opinions; the French--DeGroot model can be recovered by setting $\mu_i = 0$ for all individuals. Negative weights $w_{ij}<0$ can be used to capture antagonistic or competitive behaviors. If negative $w_{ij}$ are allowed, then setting $\mu_i = 0$ recovers the Altafini model~\cite{altafini2013antagonistic_interactions}. Hence, our model encompasses and generalizes standard models used in opinion dynamics.

A stubborn node $s\in \mathcal V$ can be introduced by setting $\mu_s = 0$ and $w_{ss}=1$ (which implies that $w_{si}=0$, for all $i\neq s$). Then, opinion of individual $s$ remains constant for all time, i.e., $y_s(t+1) = y_s(0)$ for all $t\geq 0$. 

The convergence properties of opinion dynamics models have been extensively studied and many results can be found in two review papers by Proskurnikov and Tempo ~\cite{proskurnikov2017tutorial,proskurnikov2018tutorial_2}. A key result, which will be used in the sequel, is the following.

\begin{proposition}[Theorem~$2$ from Chen \textit{et al.}~\cite{chen2019randomDeGroot}]\label{prop:stubborn}Let $W$ be such that $w_{ij}\geq 0$ for all $(i,j) \in \mathcal E_W$ and $\sum_{j=1}  w_{ij} = 1$ for all $i\in\mathcal{V}$. Suppose that there is a single stubborn node $s$ that is reachable from all other nodes on the communication layer\footnote{A node $s\in\mathcal V$ is reachable from $r\in\mathcal V$ if and only if there exists a sequence of nodes $(i_1=r,i_1,\dots, i_\ell=s)$ such that $(i_{i},i_{i+1})$ is an edge, for any $i\in\{1,\dots,\ell-1\}$.}, and let  $\mu_i=0$, for all $i\in\mathcal V\smallsetminus\{s\}$. Then, under~\eqref{eq:opinion}, $y_i(t)\to y_s(0)$ for all $i\in\mathcal V$ almost surely. That is, the opinion of every individual converges to the opinion of the stubborn node with probability $1$.
\end{proposition}

\subsection{\label{ssec:decision}Decision-Making Component}

The decision-making mechanism is developed within the framework of evolutionary game theory~\cite{smith1982book}. Specifically, each individual's action is updated according to a noisy best response~\cite{blume1995best_response} which evolves according to the log-linear learning rule in~\eqref{eq:loglinear}, regulated by the level of rationality $\beta_i\geq 0$. In the limit of no rationality, that is, $\beta_i=0$, actions are chosen uniformly at random and independent of the payoff, since $\mathbb{P}(x_i(t+1) = +1)=\mathbb{P}(x_i(t+1) = -1)=1/2$. 
The case $\beta_i=\infty$, instead, models the fully rational scenario, in which individuals always choose to maximize their payoff so that~\eqref{eq:loglinear} reduces to a deterministic best response dynamics:
\begin{equation}\label{eq:best_response}
    \mathbb{P}(x_i(t\!+\!1) \!=\! +1) =\left\{\begin{array}{ll}1&\text{if }\pi_i(+1|y_i,x_{-i})>\pi_i(-1|y_i,x_{-i}),\\\frac12&\text{if }\pi_i(+1|y_i,x_{-i})=\pi_i(-1|y_i,x_{-i}),\\0&\text{if }\pi_i(+1|y_i,x_{-i})<\pi_i(-1|y_i,x_{-i}).\\\end{array}\right.
\end{equation}
For bounded levels of rationality, $\beta\in(0,\infty)$, individuals are allowed to choose both actions, but select the one that maximizes their payoff with higher probability.

We now elucidate how different factors impact the payoff of an individual, including an individual's opinion, the actions of his or her neighbors,  the individual's commitment, and the evolutionary advantage determine his or her payoff. Observe that the payoff for taking action $+1$ and $-1$ are equal to:
\begin{subequations}\begin{align}\label{eq:payoff_spec1}
    &\pi_i(+1| y_i, x_{-i})  \displaystyle= \frac{1}{2}\lambda_iy_i + (1-\lambda_i)\frac{1}{2d_i}\sum_{j=1}^n a_{ij}(1+\alpha)(1+x_j),
      \intertext{and}
    &\pi_i(-1 | y_i, x_{-i})  \displaystyle= -\frac{1}{2}\lambda_iy_i + (1-\lambda_i)\frac{1}{2d_i}\sum_{j=1}^n a_{ij}(1-x_j),\label{eq:payoff_spec2}
 \end{align}
 \end{subequations}
 respectively. The first term accounts for the opinion $y_i$, so that individual $i$ receives an increased payoff for taking the action that individual $i$ prefers. For instance, an individual with a negative $y_i$ (that is, a preference for action $-1$) will receive a component with a negative payoff $\lambda_i y_i/2$ or positive payoff $-\lambda_i y_i/2$ for taking action $+1$ or $-1$, respectively. The second term captures the social pressure  to coordinate with neighbors. For each neighbor $j$, individual $i$ receives a positive contribution to the payoff if and only if he or she takes the same action as individual $j$. The parameter $\alpha$ models the possible evolutionary advantage for taking one of the two actions with respect to the other. In a general formulation of the model, $\alpha$ can assume any real value. Without loss of generality, in this paper we assume that, if there exists an evolutionary advantage, then action $+1$ has an evolutionary advantage with respect to $-1$, yielding $\alpha \geq 0$. The commitment $\lambda_i$ measures how much individual $i$ values and is committed to his or her own opinion during the decision-making process relative to a desire to coordinate with the neighbors' actions; setting $\lambda_i = 0$ recovers the network coordination game, which has been widely used to study  diffusion of innovation and contagion in social networks~\cite{morris2000contagion,young2011dynamics,montanari2010spread_innovation}. A stubborn node $s$ can be modeled by setting $\lambda_s=1$ and $\beta_s=\infty$, so that he or she will always take the same action $x_s (t) = x_s(0)$, for all $t\geq 0$.

By comparing~\eqref{eq:payoff_spec1} and~\eqref{eq:payoff_spec2}, we observe that the payoff for choosing action $+1$ is greater than the payoff for taking action $-1$ whenever
\begin{equation}\label{eq:threshold}
    \frac{1}{d_i} \sum_{j=1}^n a_{ij}x_j > -\frac{1}{2+\alpha} \left(\alpha+2\frac{\lambda_i}{1-\lambda_i}y_i\right),
\end{equation}
as explicitly computed in Appendix~\ref{app:threshold}. In other words, a fully rational individual $i$'s best response is action $+1$ if the above  holds. The term $\frac{1}{d_i} \sum_{j=1}^n a_{ij}x_j \in [-1, 1]$ measures the (normalized) influence on individual $i$ of the actions of his or her neighbors. Setting the commitment $\lambda_i = 0$, individual $i$ receives a higher payoff for taking action $+1$ if the influence of his or her neighbors taking $+1$ exceeds the threshold $-\alpha/(2+\alpha) \in (-1, 0]$, consistent with the results in the literature on network coordination games~\cite{young2011dynamics}. With $\lambda_i > 0$, this threshold is shifted whenever individual $i$ prefers one action over the alternative. As $y_i$ increases or decreases, the fraction of neighbors taking action $+1$ required for individual $i$'s best response to be action $+1$ decreases or increases, respectively. Thus, the proposed payoff function yields an intuitive and reasonable best-response decision-making process in which individual $i$'s threshold for selecting an action can be shaped by his or her preference for that action. Interestingly, if $\lambda_i>1/(1+y_i)$ or $\lambda_i>(\alpha+2)/(\alpha+2-y_i)$, then action $+1$ or action $-1$, respectively, always yields a better payoff than the opposite action, irrespective of his or her neighbors' current actions. In other words, if individual $i$ is simultaneously strongly committed to his or her opinion and has a strong preference for one of the two actions, then he or she will always favor that action irrespective of the social pressure.

\section{\label{sec:innovation}Adoption of advantageous innovation}

\begin{figure*}
\centering
     \subfloat[$\lambda=0.1$,  $\mu=0.001$]{\includegraphics{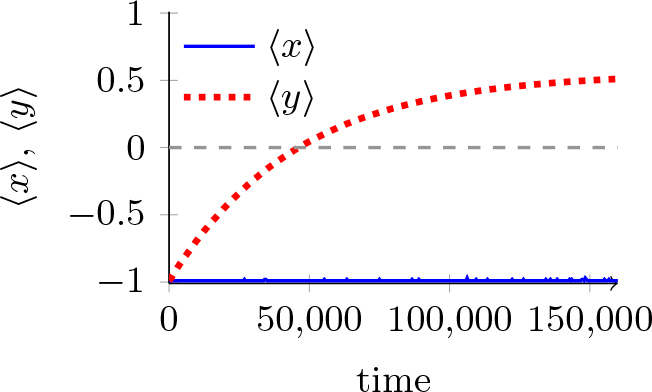}\label{fig:unpopular}}\,
 \subfloat[$\lambda=0.1$, $\mu=0.01$]{\includegraphics{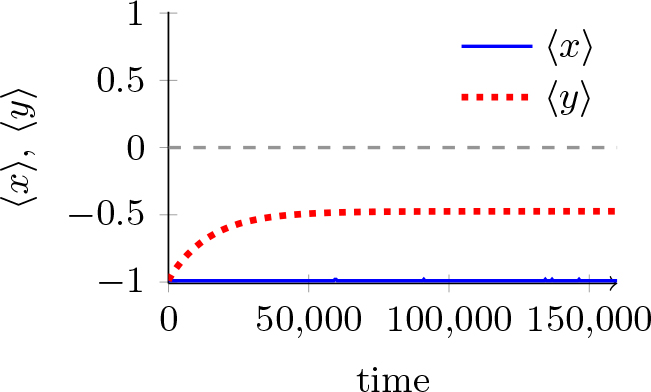}\label{fig:popular}} \,\subfloat[$\lambda=0.5$,  $\mu=0.001$]{\includegraphics{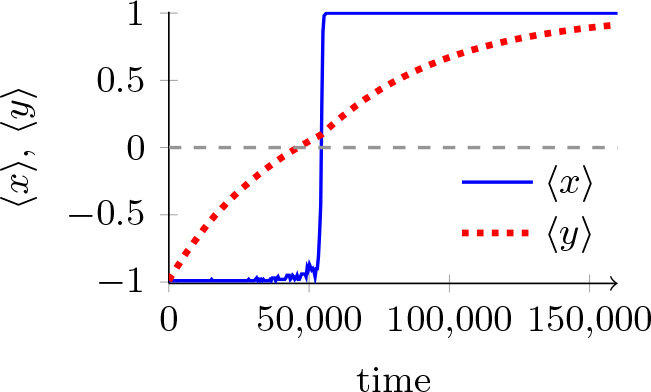}\label{fig:shift}}
    \caption{Possible outcomes of the proposed coevolutionary dynamics when modeling the diffusion of an advantageous norm. The blue solid curve is the average action of the population, $\langle x\rangle$, and the red dotted curve is the average opinion, $\langle y\rangle$. Depending on the model parameters, (a) unpopular norms, (b) popular disadvantageous norms, or (c) a paradigm shift can be observed. All sample paths are generated on networks with $n=200$ individuals, evolutionary advantage $\alpha=0.5$ and rationality $\beta=20$. Both layers are regular random graphs, with degree equal to $4$ for the communication layer and $8$ for the influence layer.  Parameters $\lambda$ and $\mu$ differ from one simulation to the other and are reported in the corresponding captions.}
    \label{fig:sample}
\end{figure*}

For the rest of this paper, we specialize the proposed framework to model and predict whether or not a social network widely adopts an advantageous innovation, and if such an adoption occurs, whether or not the innovation is actually popular among the individuals. In this section, we describe how our model is tailored to represent such a real-world process and we illustrate the different phenomena that can be typically observed as an outcome of the proposed model. In the next section, we investigate more closely the various factors that determine which phenomena is observed.

We consider a population where all the individuals start by taking the status quo (action $-1$), while one innovator $s\in\mathcal V$ is introduced in the network. The innovator is modeled as a stubborn node with fixed action and opinion equal to $x_i(t)=y_i(t)=+1$, for all $t\geq 0$, where the innovative action $+1$ has an evolutionary advantage $\alpha>0$ (see Sections~\ref{ssec:opinion} and~\ref{ssec:decision} for details on the parameters of a stubborn node).

For the sake of simplicity, we will make the following homogeneity assumptions. We assume that all the nonstubborn individuals have the same level of rationality, commitment, and susceptibility, that is,  $\beta_i=\beta$,  $\lambda_i=\lambda$, and $\mu_i=\mu$, for all $i\in \mathcal V\smallsetminus\{s\}$. We further assume that the communication layer is connected\footnote{A layer is connected if every node is reachable from all the others.} and that $W$ is a simple random walk on the communication layer, that is, the nonzero entries of any row of $W$ are all positive and of equal value, and sum to $1$. I.e., we are considering a specific implementation of a French--DeGroot model. This implies that in updating opinion $y_i$, individual $i$ gives the same weight to the opinion of each one of his or her neighbors on the communication layer.

The goal of our study is to explore the role of the coupling between the opinion dynamics and the decision-making mechanisms --- determined by the commitment $\lambda$ and the susceptibility $\mu$ --- and of the network structure on the emerging behavior of the system. To help elucidate this goal, we define the following two quantities:
\begin{equation}\label{eq:average}
\langle x\rangle:=\frac1n \sum_{i=1}^n x_i,\quad\text{and}\quad \langle y\rangle:=\frac1n \sum_{i=1}^n y_i,
\end{equation}
which are the average action and opinion in the population, respectively. In Fig.~\ref{fig:sample}, we offer three paradigmatic sample paths of the coevolutionary dynamics at the population level, exhibiting the different phenomena that can occur, which we now describe in further detail.

\begin{description}
\item[Unpopular norm (Fig.~\ref{fig:unpopular})] after a short transient, the average of the individuals' opinions  shows a preference for the innovation, that is, $\langle y\rangle>0$. However, an overwhelming majority of the individuals still takes the status quo action, that is, $\langle x\rangle\approx-1$. While the ergodic nature of~\eqref{eq:loglinear} ensures that the innovation will eventually diffuse across the entire network, we will see that the unpopular norm may be meta-stable for a long period of time. In the real world, this would imply that the widespread adoption of the innovation fails to occur.
\item[Popular disadvantageous norm (Fig.~\ref{fig:popular})] the status quo (which is disadvantageous with respect to the innovation) remains the predominant action in the network ($\langle y\rangle\approx -1$) and it is on average the preferred action among the individuals' opinions  ($\langle y\rangle<0$), for any reasonably long period of time. It is worth noticing that the comment on the ergodicity of~\eqref{eq:loglinear} for unpopular norms also applies to this case. 
\item[Paradigm shift (Fig.~\ref{fig:shift})] after a short transient, a tipping-point is reached and the advantageous innovation is adopted and supported by almost the entire population ($\langle y\rangle\approx +1$ and $\langle y\rangle>0$). It is worth noting that the spreading of the innovation is often so fast that the changes in actions occur faster than change in opinions, i.e., shortly after the tipping point, we observe $\langle x \rangle \geq \langle y \rangle$.
\end{description}

A fourth phenomenon should be in principle possible, being the establishment of a meta-stable state in which the advantageous innovation is widely adopted as the norm, but is unpopular. However, this was never observed in our numerical simulations. An intuitive reason can be found in the following consideration. If a large majority of the individuals adopt the innovation, then their opinion will drift toward $+1$ due to both the influence of the neighbors' actions and the stubborn node, thereby eventually leading to a paradigm shift.

\section{Effect of the network structure on the adoption of innovation}\label{sec:results}

In this section, we aim to understand how the model parameters and  the network structure may determine the emergence of one of the three different collective phenomena  described in Section~\ref{sec:innovation}, during the adoption of an advantageous innovation. 

We begin our analysis by considering the limit case of fully rational individuals, that is, $\beta=\infty$. We again consider the case in which a single stubborn node (termed the innovator) $s\in\mathcal V$ is introduced in the network, taking a fixed action and having opinion equal to $x_i(t)=y_i(t)=+1$, for all $t\geq 0$. We further assume that the network is connected. In this scenario, the following result can be established, with the proof found in Appendix~\ref{app:proof}.

\begin{theorem}\label{theorem}Let us consider a coevolutionary dynamics of opinions and decisions. Let us define
\begin{equation}\label{eq:dstar}
d^*:=\min\{d_i:i\in\mathcal V, (i,s)\in\mathcal E_A\}.
    \end{equation}
In the limit $\beta=\infty$, if $\alpha<d^*-2$ and
\begin{equation}\label{eq:lambdastar}
\lambda<\lambda^*:=\frac12-\frac{2+\alpha}{4d^*-4-2\alpha},
\end{equation}
then  $\langle x(t)\rangle=-1+2/n$, for all $t\geq 0$. That is, a paradigm shift cannot occur.
\end{theorem}

Theorem~\ref{theorem} yields a necessary condition for a paradigm shift; if either the evolutionary advantage or the commitment in the decision-making process is sufficiently large ($\alpha\geq d^*-2$ or $\lambda>\lambda^*$), then a paradigm shift is possible. Note that both conditions depend on the network structure through the minimum degree of the neighbors of the innovator $d^*$. 

Such conditions are not sufficient, however. In fact, one can easily produce simple examples in which even though the conditions in Theorem~\ref{theorem} are satisfied, a paradigm shift does not occur since the diffusion of action $+1$ might stop after a few adoptions (for instance, if there is a bottleneck in both layers of the network). Further analysis, envisaged as future research, is required to establish sufficient conditions for diffusion, which are likely to depend on the overall structure of both layers of the network, and not only on the nodes directly connected to the innovator on the influence layer.

\begin{figure*}
    \centering
     \subfloat[RR, average decision]{   \label{fig:sample2}
\includegraphics{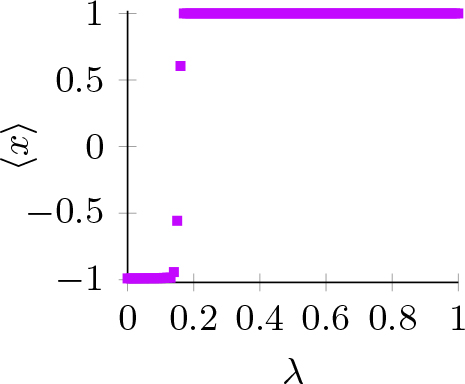}} \,\,
  \subfloat[ER, average decision]{\includegraphics{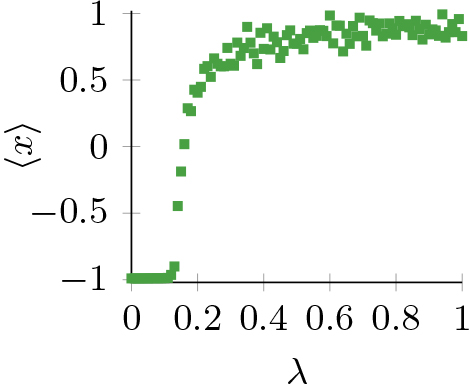}} \,\,
 \subfloat[WS, average decision]{\includegraphics{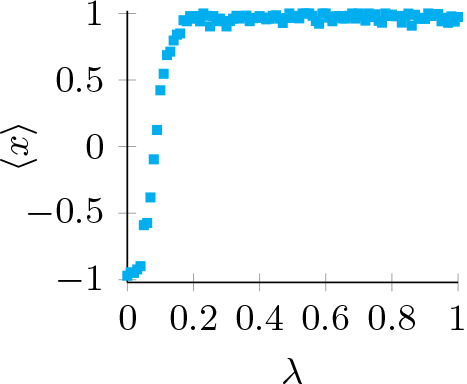}}\,\, \subfloat[BA, average decision]{\includegraphics{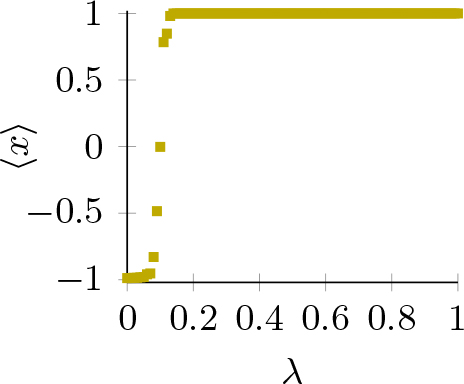}}\\
 \subfloat[RR, variance]{\includegraphics{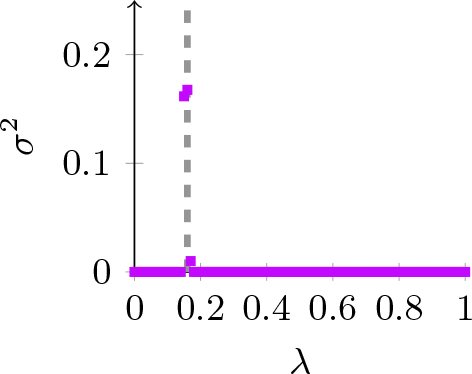}} \,\,
  \subfloat[ER, variance]{\includegraphics{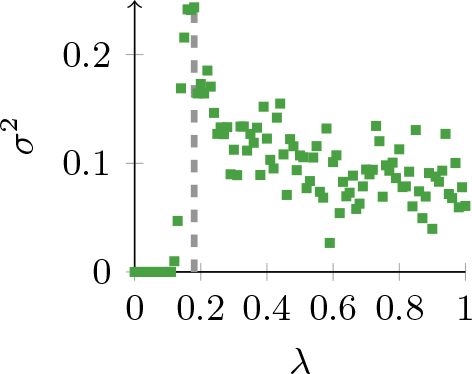}} \,\,
 \subfloat[WS, variance]{\includegraphics{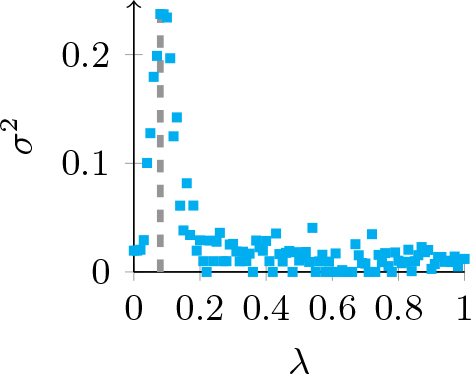}} \,\,\subfloat[BA, variance]{\includegraphics{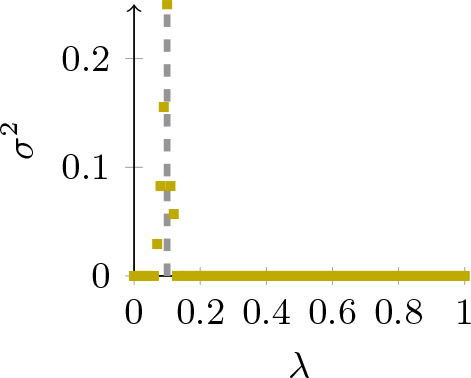}}
    \caption{Estimation of the threshold value for $\lambda$ for transitioning from meta-stable unpopular norm to paradigm shift. Average action of the population (in (a)--(d)) and variance in the fraction of $+1$ actions (in (e)--(h))} at time $T=4n^2$ over $100$ independent runs of the coevolutionary dynamics with $\alpha=0.5$, $n=200$, $\beta=20$, and both layers with average degree $8$ (influence layer) and $4$ (communication layer), generated according to (a) a random regular graph, (b) an Erd\H{o}s-R\'enyi random graph, (c) a Watts-Strogatz small-world network with rewiring probability $p=0.2$, and (d) a Barab\'asi-Albert scale-free network.
    \label{fig:zealot}
\end{figure*}

In the rest of this section, we will instead focus on the scenario in which individuals have a bounded level of rationality, that is, $\beta<\infty$, which has been demonstrated to be more consistent with real-world decision-making processes~\cite{simon2000bounded}. In this case, we will see that paradigm shifts may occur even for levels of commitment $\lambda<\lambda^*$. In order to focus on the effect of the coupling between the two mechanisms and of the network on the system's evolution, the following numerical studies will fix a moderate level of evolutionary advantage $\alpha=0.5$ and a sufficiently large level of rationality $\beta=20$, while studying the behavior of the system for different values of $\lambda$ and $\mu$, and for different network topologies. These parameters aim to capture a realistic scenario in which the evolutionary advantage of the innovation action $+1$ is present, but does not have such a dominant role as to make the contributions of the other dynamics negligible, and individuals with bounded rationality still maximize their payoff  with sufficiently high probability (for instance, the probability of deviating from a fully established and supported norm is less than $10^{-11}\%$). The quantitative results of the following numerical simulations may depend on the precise choice of the parameters $\alpha$ and $\beta$. However, we have observed that the salient features of the observed phenomena of the system are robust to different choices of the parameters $\alpha$ and $\beta$ that represent the described scenario.

\subsection{Opinions not directly influenced by actions}\label{ssec:mu0}

In the first part of our analysis, we will assume that the evolution of the opinion is not influenced by the actions, i.e., with susceptibility $\mu=0$. Since we assume that the communication layer is connected,  Proposition~\ref{prop:stubborn} establishes that the opinions of all individuals converge almost surely to $+1$. Hence, only two phenomena can occur: unpopular norm or paradigm shift. Before starting our analysis for bounded rational individuals, we briefly report a straightforward consequence of Theorem~\ref{theorem} and Proposition~\ref{prop:stubborn} for the behavior of the coevolutionary dynamics with fully rational individuals when $\mu=0$.

\begin{corollary}\label{corollary}Let us consider a coevolutionary dynamics of opinions and decisions with $\mu=0$. Let $W$ be such that $w_{ij}\geq 0$ for all $(i,j)$ in $\mathcal E_W$, $\sum_{j=1}  w_{ij} = 1$ for all $i\in\mathcal{V}$. In the limit $\beta=\infty$, if $d^*>2+\alpha$ and $\lambda<\lambda^*$, then $\langle x(t)\rangle=-1+2/n$, for all $t\geq 0$, and $\langle y(t)\rangle\to 1$. That is, a paradigm shift cannot occur, and rather, an unpopular norm is almost surely observed.
\end{corollary}

One can intuitively conjecture that, if opinions play a sufficiently dominant role in the decision-making process (that is, $\lambda$ is sufficiently large), then the whole network will adopt the innovation, while, in the opposite scenario, the social pressure outweighs the individual's commitment to his or her own opinion, thus ensuring the population continues to choose the status quo action, even though the opinion of the overwhelming majority shows preference for the innovation. Indeed, evidence of a phase transition depending on the commitment $\lambda$ can be observed in Fig.~\ref{fig:sample2}.

We investigate the presence of such a phase transition by means of Monte Carlo numerical simulations, following a method similar to the ones proposed to numerically estimate the epidemic threshold in epidemic models~\cite{moinet2018epidemics,zino2018memory}. Specifically, we run repeated independent simulations of the process for different values of commitment $\lambda$, keeping track of the fraction of adopters of the innovation in each run at the end of a fixed observation window of duration $T$, which is equal to $(\langle x(T) \rangle+1)/2$. Then, the threshold $\hat \lambda$ is estimated as the value of $\lambda$ that maximizes the variance of such a quantity within the independent runs. Sharp peaks of the variance are evidence of an explosive phase transition between a regime where unpopular norms are meta-stable (if $\lambda<\hat\lambda$), to a regime where paradigm shift is observed in almost all the simulations (if $\lambda>\hat\lambda)$. We fix a sufficiently long time-window $T=4n^2$ (each individual thus revises his or her action and opinion on average $4n$ times) to allow the innovator to steer the whole population to an opinion close to $+1$.  If an unpopular norm persists even after $T=4n^2$, then it is meta-stable, and implies that the innovation will never realistically be adopted in the real world.

To better elucidate the role of the network topology in determining such a threshold, we apply the Monte Carlo-based technique on four classical network models with different features~\cite{newman2010networks_book}. Specifically, we considered random regular (RR) graphs,  Erd\H{o}s-R\'enyi (ER) random graphs (which have a slight heterogeneous degree distribution), Watts-Strogatz (WS) small-world networks (which are characterized by a clustered structure), and Barab\'asi-Albert (BA) scale-free networks (which have a strongly heterogeneous degree distribution with a few hubs with high degree). In all our simulations, both layers of the network are generated according to the same network model, one independent of the other, and the innovator is placed in the first node, that is, $s=1$. In the case of BA networks, this would imply that the innovator is almost surely placed in a hub. In order to avoid possible confounding due to the network density when comparing different network topologies, we keep the average degree to be the same between different network structures: in the influence layer all networks have average degree equal to $8$ and $4$ in the influence layer and communication layer, respectively. More details on the generation of the networks can be found in Appendix~\ref{app:network}.

The results of our Monte Carlo simulations, presented in Fig.~\ref{fig:zealot}, confirm our conjecture  and suggest the presence of a phase transition, which is estimated to occur at the value $\hat\lambda$ indicated by the peak (vertical dashed line). When comparing the numerical estimations in Fig.~\ref{fig:zealot} with the necessary condition to achieve paradigm shift from Corollary~\ref{corollary}, it appears that bounded rationality favors the emergence and establishment of the innovation, thereby leading to a paradigm shift for values of commitment $\lambda$ that are smaller than the necessary value $\lambda^*$ for fully rational individuals. In fact, for RR (where $d^*=8$), we compute $\lambda^*=0.4074$, while the  threshold estimated numerically is $\hat\lambda=0.16$. Similarly, for the BA (where, by construction, $d^*\geq 4$), we obtain  $\lambda^*\geq 0.2727$, while the  threshold estimated numerically is $\hat \lambda=0.1$. For the the other two cases (ER and WS), $\lambda^*$ depends on the specific realization of the network, since the degrees are nonuniform and random variables. However, using their expected values, we obtain $\mathbb E[\lambda^*]=0.3234$ and $\mathbb E[\lambda^*]=0.3967$ for ER and WS, respectively. Our numerical simulations, instead, suggest that the threshold for the ER graph is equal to $\hat \lambda=0.18$, while the one for WS is estimated as $\hat \lambda=0.08$ and seems to vanish (since even for $\lambda=0$ some simulations show a paradigm shift occurring). Among the network structures, WS and AB networks seem to especially favor paradigm shifts; possible reasons can be found in the high level of clustering that characterizes WS networks (see below for further discussions) and because most of the low-degree nodes in BA networks are connected to the innovator, who is almost always positioned in a hub.

\begin{figure}
    \centering
\includegraphics{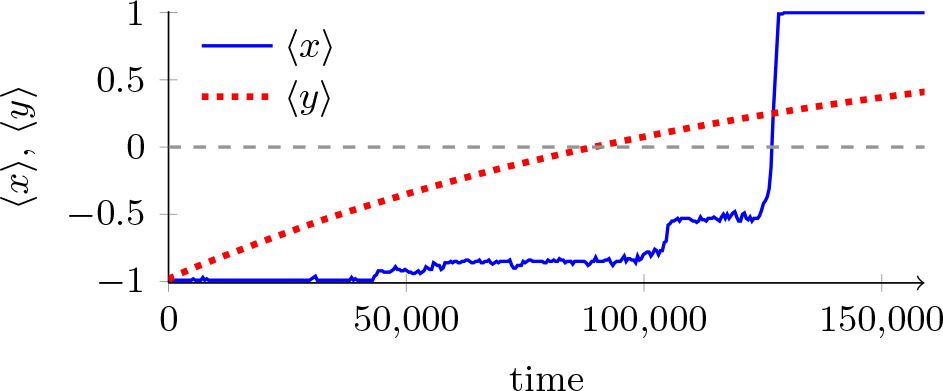}
    \caption{Average action and opinion of the population in a sample path of the coevolutionary dynamics on Watts-Strogatz small world networks with rewiring probability $p=0.2$. Other parameters are $\alpha=0.5$, $n=200$, $\beta=20$, $\lambda=0.1$, average degree $8$ in the influence layer and $4$ in the  communication layer.}
    \label{fig:WS}
\end{figure}

The results further suggest that, apart from determining the threshold value $\bar \lambda$, the network structure plays an important role in shaping the phase transition. Notice that in RR and AB networks (Figs.~\ref{fig:zealot}(a) and (d)), the phase transition seems to be extremely sharp: if $\lambda$ is below the threshold, then an unpopular norm is observed in almost all the simulations, while above the threshold, a paradigm shift is almost always observed. On the contrary, in ER and WS networks (Figs.~\ref{fig:zealot}(b) and (c)), the threshold seems to be less sharp as $\lambda$ increases, suggesting the existence of a region for the commitment $\lambda$ where both unpopular norms and paradigm shifts are possible, depending on the specific realization of the network. We believe that such a phenomenon might be caused by the variability in the degree of the innovator, which in ER and WS networks depends on the specific realization. In contrast, all the nodes in RR networks have the same degree, and the innovator is almost always placed in a hub in BA networks, by construction. For WS networks, we also observe that the region of the parameter space in which paradigm shifts can never occur seems to vanish. This may be due to the high levels of clustering in small-world networks, which helps the spread of innovation~\cite{montanari2010spread_innovation}. In fact, the sample path in Fig.~\ref{fig:WS} of a WS network shows an interesting transient phenomenon; an increasing nonzero fraction of the population adopts the innovation in steps, and persists in the adoption even though remaining in the minority. We conjecture that this occurs because the high clustering structure results in certain clusters where the individuals have mostly adopted the innovation and remain meta-stable, while in other clusters, the status quo is still widely adopted.

\begin{figure*}
    \centering
 \subfloat[RR]{\includegraphics{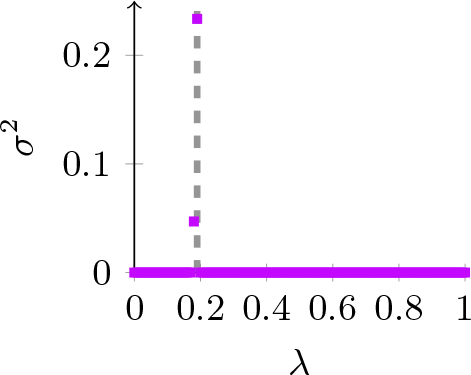}} \,\,
  \subfloat[ER]{\includegraphics{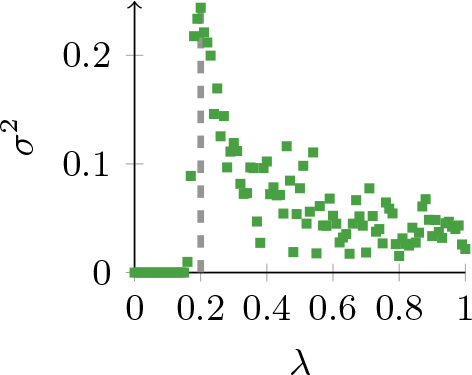}} \,\,
 \subfloat[WS ($p=0.2$)]{\includegraphics{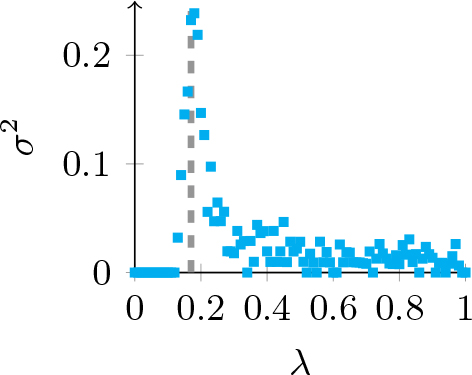}} \,\,\subfloat[BA]{\includegraphics{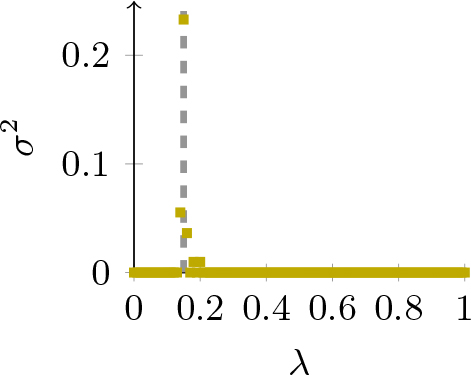}}
    \caption{Estimation of the threshold for different network topologies: (a) regular random, (b) Erd\H{o}s-R\'enyi, (c) Watts-Strogatz, and (d) Barab\'asi-Albert. Each data point is the variance of the fraction of $+1$ actions at time $T=4n^2$ over $100$ independent runs of the coevolutionary dynamics with $\alpha=0.5$, $n=200$, $\beta=20$, and both layers with average degree $16$ (for the influence layer) and $4$ (for the communication layer), generated with the four network models denoted in the corresponding caption.}
    \label{fig:dense}
\end{figure*}

Our theoretical findings in Corollary~\ref{corollary} for fully rational individuals suggest that the density of the influence layer has a detrimental effect on the diffusion process, hindering the emergence of a paradigm shift. In fact, the threshold $\lambda^*$ increases as $d^*$ increases, and approaches $1/2$. For bounded rational individuals, we investigate the effect of the density of the influence layer by repeating the numerical estimation of the threshold $\hat\lambda$ performed in the above, doubling the average degree of the influence layer to $16$.

Consistent with the intuition coming from our analytical result in the limit of fully-rational individuals, the results of our numerical study, reported in Fig.~\ref{fig:dense}, indicate that denser networks lead to an increased threshold $\hat\lambda$. However, the magnitude of such an increase seems to be strongly dependent on the network topology. For ER and RR networks, the increase is quite moderate ($10\%$ and $18.75\%$, respectively), whereas BA and WS networks seem impacted more significantly by the network density increase, yielding an estimated threshold increased of $50\%$ and $112.5\%$, respectively. Interestingly, the detrimental effect of increasing the network density is stronger for those networks in which paradigm shifts are favored on sparser networks, decreasing the differences between the four network structures. This suggests that the beneficial effect of clustering and of having the innovator placed in the hub is reduced as the network becomes denser. The case of WS networks is also of particular interest since, differently from the sparser scenario, a vanishing threshold is not observed here: if $\lambda$ is small, then unpopular norms are observed in almost all the simulations.

Depending on the communication layer topology, convergence of the individuals' opinions to $+1$ may be extremely slow, consequently hindering convergence of actions in a reasonable time-window~\cite{proskurnikov2017tutorial,proskurnikov2018tutorial_2}. To sum up, the topology of both layers may be key in predicting whether the spread of an innovation will fail, even though it has an evolutionary advantage with respect to the status quo, and further, even in scenarios where the majority of the population's opinions favor it. 

\subsection{Feedback between opinion and actions}\label{ssec:feedback}

In the previous section we have extensively analyzed the limit case of $\mu=0$, in which the actions of an individual's neighbors do not directly influence the opinion dynamics~\eqref{eq:opinion}. However, this assumption may be overly simplistic in real-world scenarios, where evidence of such an influence has been theorized in the social-psychological literature~\cite{haidt2001emotional, gavrilets2017interiorization} and observed in empirical studies~\cite{lindstrom2018role}. 

In this section, we will study the general case of susceptibility $\mu>0$. In this case, we immediately notice that the innovator is not necessarily always able to steer the opinions of all individuals to $+1$. As a consequence, all three phenomena reported in Section~\ref{sec:innovation} and illustrated in Fig.~\ref{fig:sample} can be observed depending on the value of the commitment $\lambda$ and susceptibility $\mu$. In particular, we will now explore in detail the interplay between susceptibility and commitment, and the role of the network structure in determining the outcome of the coevolutionary dynamics. Specifically, we choose the two topologies analyzed in the previous subsection that produced the greatest differences in observed outcomes, i.e., the RR and WS networks. For each one of these topologies, we estimate the average opinion and action of the population at time $T=4n^2$ by means of $100$ independent simulations, while varying the values of both parameters $\lambda$ and $\mu$.

\begin{figure*}
    \centering
 \subfloat[RR network: average action $\langle x \rangle$]{\includegraphics{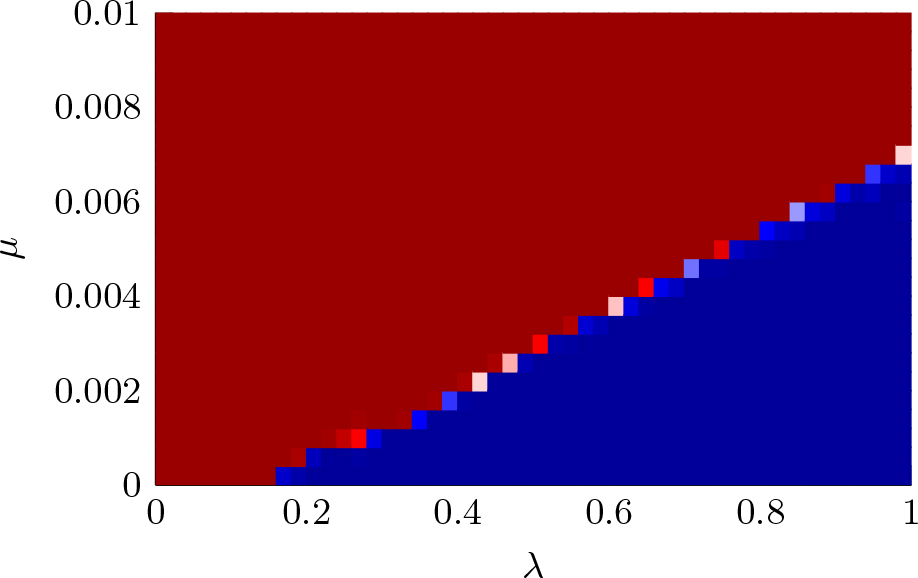}}
  \subfloat[WS network: average action $\langle x \rangle$]{\includegraphics{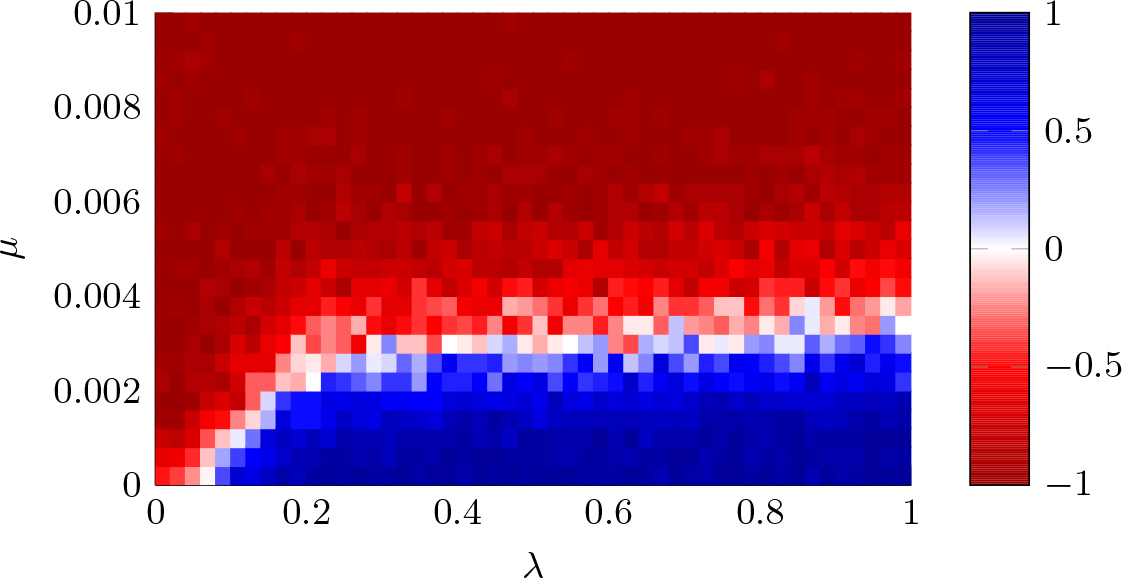}}\\ \subfloat[RR network:  average opinion $\langle y \rangle$]{\includegraphics{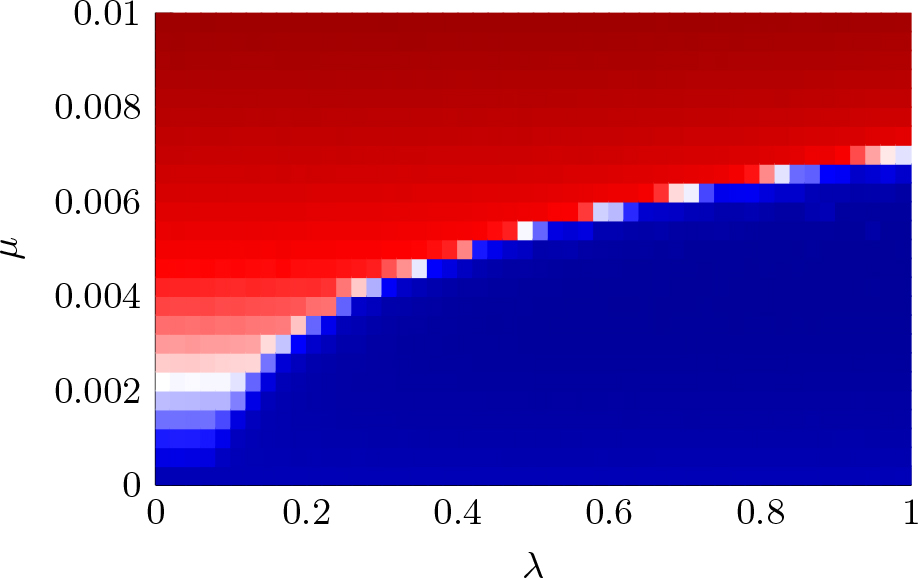}}
  \subfloat[WS network: average opinion $\langle y \rangle$]{\includegraphics{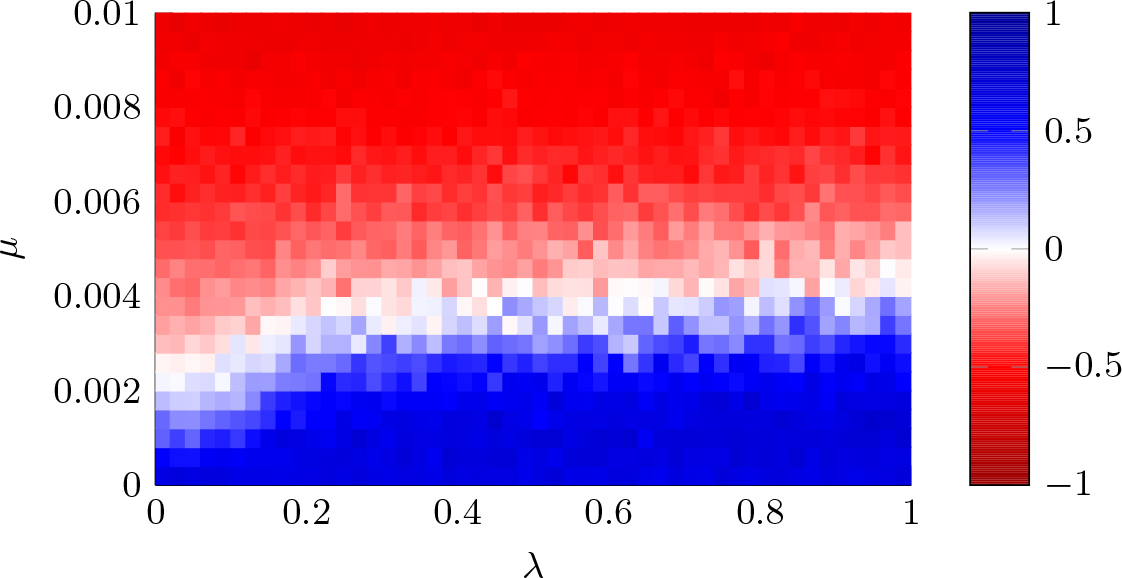}}
  \\ \subfloat[RR network:  regions]{\includegraphics{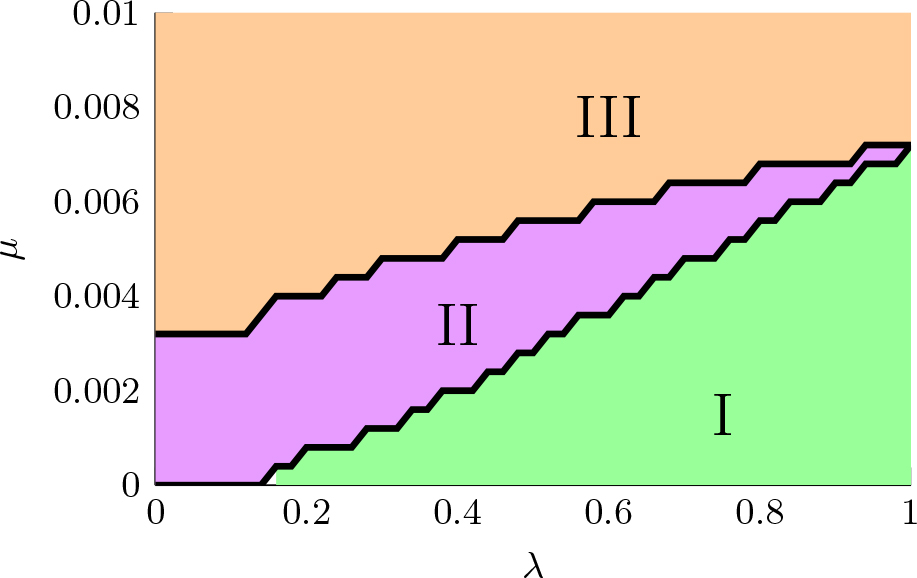}}
  \subfloat[WS network: regions]{\includegraphics{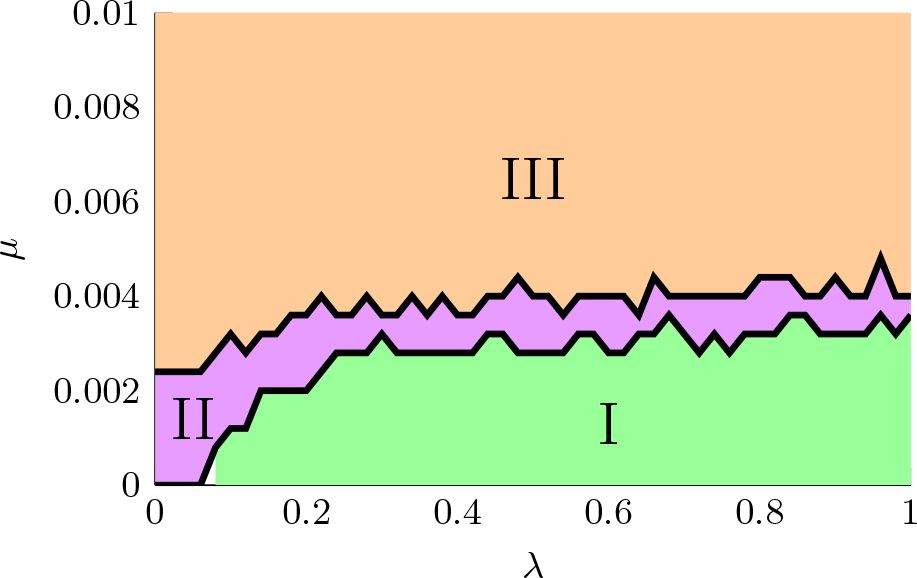}\qquad\qquad\quad}
    \caption{Outcome of coevolutionary dynamics on (a,c,e) RR and (b,d,f) WS graphs for different values of the parameters $\lambda$ and $\mu$. The average action is plotted in (a--b), the average opinion in (c--d), and in (e--f) we highlight the three distinct regions of the parameter space associated with the emerging phenomenon observed in the simulations. Each data point is the average over $100$ independent runs with $\alpha=0.5$, $n=200$, $\beta=20$, over a time-window of duration $T=4n^2$. Both types of graphs have average degree $8$ (for the influence layer) and $4$ (for the communication layer). The rewiring probability of the WS networks is $p=0.2$. The green region I, violet region II, and orange region III correspond to regions in which a paradigm shift, an unpopular norm, and a popular disadvantageous norm, is observed, respectively.}
    \label{fig:feedback}
\end{figure*}

The results of our numerical simulations are reported in Fig.~\ref{fig:feedback}. Comparing the average action in (a) and (c) with the corresponding average opinion in (b) and (d), we identify three regions corresponding to the three possible phenomena that can occur, highlighted in (e) and (f), respectively. In the green region denoted by the roman number I, the commitment $\lambda$ is sufficiently large and the susceptibility $\mu$ is small, and we thus observe  paradigm shifts ($\langle x(T) \rangle\approx +1$ and $\langle y(T) \rangle>0$). In the violet region, denoted as II, we observe the emergence of unpopular norms, whereby $\langle x(T) \rangle\approx -1$ and $\langle y(T) \rangle>0$. For higher levels of susceptibility, we finally find region III (in orange), in which $\langle x(T) \rangle\approx -1$ and $\langle y(T) \rangle<0$, signifying the presence of meta-stable disadvantageous popular norms.

The shape of these regions and the sharpness of the phase transition between each region is strongly influenced by the network structure. In RR networks (Figs.~\ref{fig:feedback} (a), (c), and (e)), we mostly observe sharp phase transitions between the regimes. In particular, a popular disadvantageous norm is almost always observed if $\mu>0.007$ (region III), regardless of the commitment $\lambda$. For intermediate values of susceptibility, that is, $0.0032<\mu<0.007$, there are instead two phase transitions, at two different values of commitment, denoted as $\lambda'<\lambda''$. Specifically, if $\lambda<\lambda'$, then we observe the emergence of a disadvantageous popular norm (region III); for $\lambda'<\lambda<\lambda''$, we have an unpopular norm (region II); if $\lambda>\lambda''$, a paradigm shift is observed (region I). Finally, if the susceptibility is small, that is, $\mu<0.0032$, we recover the findings in Section~\ref{ssec:mu0}, where depending on $\lambda$, we observe either an unpopular norm (region II) or a paradigm shift (region I).

In the case of WS networks (Figs.~\ref{fig:feedback} (b), (d), and (f)), we immediately observe that all the phase transitions appear to be less sharp, similar to what was already reported in Section~\ref{ssec:mu0} on the case of $\mu=0$. The three regions described above also appear to have a different shape with respect to those observed in RR networks. In fact, if the commitment $\lambda>0.25$, then the precise value of $\lambda$ seems not to play any role in determining the outcome of the process, and rather, the outcome is instead uniquely determined by the susceptibility: if $\mu<0.003$, then paradigm shifts occur (region I); if $0.003<\mu<0.004$, we observe the emergence of unpopular norms (region II); finally, for $\mu>0.004$, popular disadvantageous norms persist (region III). For $\lambda<0.25$, the behavior is similar to the one already described for RR networks.

When comparing the two topologies, we can conclude that the introduction of a direct feedback of the observed actions on an individual's opinion has a different effect, depending on the network structure. For instance, topologies that seems to favor the occurrence of paradigm shifts in the absence of such a feedback (e.g., WS netwoks), are instead less prone to promote the diffusion of innovation when the feedback is present, thereby favoring the emergence of popular disadvantageous norms. We believe that the presence of clustering in WS networks may explain this phenomenon. In fact, as $\mu$ increases, the ability of the innovator to shift others' opinions may remain restricted to individuals in his or her own immediate cluster, while in the other clusters at a longer path distance from the innovator, the influence of observed actions on an individual's opinion may ensure that the majority of the individuals' opinions remains firmly in support of the status quo action.

\section{\label{sec:conclusion}Discussions and Conclusions}

In this paper, we have proposed a novel modeling framework for capturing the intertwined coevolution of individuals' opinions and their actions;  individuals share their opinions and are influenced by the actions observed from the other members of their community on two distinct layers of a complex social network. The first key contribution of this work is the formal definition of the coevolutionary model itself, which is grounded in, and intertwines, the theories of opinion dynamics and evolutionary games.  

Then, we have tailored the proposed framework to study a real-world application, concerning the introduction of an innovation (such as a novel advantageous product or behavior) in a social community. In Section~\ref{sec:innovation}, we have provided details of such a model, illustrating three very different real-world phenomena that can be observed within our single unified modeling framework: the formation of i) an unpopular norm, ii) a popular disadvantageous norm, or iii) a paradigm shift. 
The possible formation of either unpopular norms or popular disadvantageous norms has been rarely considered in agent-based models, even though substantial empirical data and studies from the social-psychology literature indicate neither phenomenon is especially rare~\cite{abbink2017_badsocialnorms,prentice1993pluralistic_alcohol,willer2009false_enforcement,mackie1996footbinding,Smerdon2019}. If indeed a paradigm shift does occur, it interesting that the opinions are first to change, followed by the actions. In a real-world example of such a phenomenon, Iowa farmers in the 1930s began widespread adoption of a new hybrid corn; it was during the prior years that farmers gradually learned about the new hybrid corn and slowly shifted their opinion toward supporting its adoption~\cite{ryan1943diffusion_corn}. 

A preliminary analysis established a necessary condition on the model parameters and network structure to observe a paradigm shift when individuals actions are fully rational. However, real-life human cognitive processes have been demonstrated to be only partially rational~\cite{simon2000bounded}, and the case of bounded rationality was then studied by means of Monte Carlo numerical simulations. We started from a simplified scenario, in which individuals' opinions are not susceptible to the observed actions of the neighbors. Evidence of a phase transition between two regimes of a meta-stable unpopular norm and a paradigm shift was identified, based on the strength of individuals' commitment to their own opinion, and further shaped by the network structure. This accords with intuition: if an individual's decision-making is primarily governed by the desire to coordinate with his or her neighbors, then it becomes unlikely that the social system collectively breaks out of the meta-stable state in which individuals largely select the status quo action, even though the innovator may shift the opinions of the community to support the innovation. 

Our analysis was then extended to consider the more realistic scenario in which an individual's opinion is susceptible to the influence of the  observed actions of others in their social community. The results illustrated that the three social phenomena mentioned above could all be observed, depending on the model parameters. The range of parameter values for which each phenomena could occur as well as the sharpness of the phase transition between the different regimes was found to be strongly dependent on the topology of the social network. An important general conclusion was also drawn. If individuals' opinions are strongly susceptible to being influenced by the actions of others, then independent of the network topology and of the individuals' commitment to their own opinion, the status quo will persist as a popular disadvantageous norm. The model can thus shed light on why some norms persist even though they are clearly disadvantageous to both the individual and the wider population. For instance, footbinding was a disadvantageous norm among Chinese women for several centuries prior to a rapid disappearance in the 20th Century, persisting in part because individuals' opinions were heavily influenced by the observed actions of others~\cite{mackie1996footbinding}.

In contrast, having a large commitment to one's own opinion is a  necessary, but not sufficient, condition to observe a paradigm shift (see Fig.~\ref{fig:feedback}). This illustrates the importance of individuality, or the role of an individual's evolving preference for/opinion on an action, in promoting the spread of an innovation. Such a role, despite being intuitive, has been largely overlooked in most diffusion literature. For all topologies, the range of parameter values for which an unpopular norm could occur was nontrivial (region II in Fig.~\ref{fig:feedback}). This helps support the observations from the literature that unpopular norms, while not extremely common, are also not rare. The high clustering nature of small-world networks has been linked to paradigm shifts occurring rapidly when considering just a decision-making process~\cite{montanari2010spread_innovation}. In the coevolutionary model, we found that if an individual's susceptibility to having his or her opinion influenced by the observed actions of others is small, then small-world networks favor the adoption of novel advantageous norms, consistent with~\cite{montanari2010spread_innovation}. However, as the susceptibility to social influence increases, small-world networks become more resistant to the diffusion of innovation than other network structures (e.g., random regular graphs), thus leading to the emergence of popular disadvantageous norms. Thus, we confirm that the network structure itself plays a non-negligible role in shaping the collective dynamics, but when decision-making and opinion dynamic co-evolve, the impact can be unexpected and counter-intuitive.

We hope to have convinced the reader that the proposed coevolutionary modeling framework is of interest to the various scientific communities that study social systems using dynamical mathematical models. The general formulation of our modeling framework and the promising preliminary results obtained have paved the way for several avenues of future research. On the one hand, further efforts should be devoted toward a rigorous theoretical analysis of the model, beginning with a comprehensive convergence result for the fully rational case. Further analysis of the network topology may be considered, including the impact of introducing directed interactions on either layer, or the effect of negative interaction weights on the emergence of polarization phenomena~\cite{altafini2013antagonistic_interactions}. The role of clustering in favoring or hindering the occurrence of a paradigm shift should also be investigated, especially in the presence of strongly connected components in directed networks or communities with negative weights. Time-varying networks are recognized as being more realistic, with several possible directions including activity-driven networks~\cite{zino2019opinion}, adaptive topologies~\cite{Holme2006coevolution}, or state-dependent weights~\cite{lorenz2007opinion}. Moreover, while a coordination game was used for the decision-making process, the proposed framework can easily be adjusted to consider other network games, such as anti-coordination, Prisoner's Dilemma, etc. Among the several fields in which the proposed framework may find application, we want to mention marketing and financial markets. In marketing and product promotion, it has often been observed that the mere fact that a novel product is superior to the competitors may be not sufficient for it to succeed, even if the superiority is widely acknowledged. The proposed framework can offer mathematical tools to represent realistic diffusion of a new product and predict its outcome. 

Existing literature~\cite{curty2006phase,da2015sudden,delellis2017evolving} has recognized that in financial markets, there is a coevolution of a trader's (individual) reputation and trading strategies; while the reputation can be generally modeled through a continuous variable (similar to opinions), the trading strategies can either be represented as edge creation/deletion operations~\cite{da2015sudden}, or as a complex decision-making process~\cite{delellis2017evolving}. Ideas drawn from these existing works may enable our proposed framework to better describe the phenomena studied in this work and suggest their possible extension to the particular application of financial markets.

\section*{Data Availability Statement}
The data that support the findings of this study are available from the corresponding author upon reasonable request.

\begin{acknowledgments}
This work was partially supported by the European Research Council (ERC-CoG-771687) and the Netherlands Organization for Scientific Research (NWO-vidi-14134). M. Ye was also partially supported by Optus Business.
\end{acknowledgments}

\appendix
\section{Proof of Proposition~\ref{prop:well}}\label{app:opinion}

We prove that if $y_i(0)\in[-1,1]$ for all $i\in\mathcal V$, then $y_i(t)\in[-1,1]$, for all $i\in\mathcal V$ and for all $t\geq 0$. That is, the opinions are always well defined if the initial opinions are well defined. We proceed by induction. Let us assume that the opinions are well posed at a generic time $t$, that is, $y_i(t)\in[-1,1]$, for all $i\in\mathcal{V}$. Let $i \in \mathcal{V}$ be the individual that is activated at time $t$. Clearly, all the opinions of the individuals $j\in\mathcal V\smallsetminus\{i\}$ are well defined, since they remain the same. For the opinion of individual $i$, we obtain the following bound: 
\begin{equation}\begin{array}{lll}
  |y_i(t+1)| &=&\displaystyle \left|(1-\mu_i)\sum_{j=1}^n w_{ij}y_j(t)+ \mu_i \frac{1}{d_i} \sum_{k=1}^n a_{ik} x_k(t)\right|\\
  &\leq&\displaystyle(1-\mu_i)\left|\sum_{j=1}^n w_{ij}y_j(t)\right|+ \mu_i \left|\frac{1}{d_i} \sum_{k=1}^n a_{ik}x_k(t)\right|\\
    &\leq&\displaystyle(1-\mu_i)\sum_{j=1}^n |w_{ij}||y_j(t)|+ \mu_i \frac{1}{d_i} \sum_{k=1}^n a_{ik}|x_k(t)|\\
&\leq&\displaystyle(1-\mu_i)\sum_{j=1}^n |w_{ij}|+ \mu_i\leq 1.
    \end{array}\end{equation}
Since we have selected a generic time $t$ and the bound holds for all $i\in \mathcal{V}$, the opinions at the next time step are well defined: $y_i(t+1) \in [-1, 1]$. This yields the proof.

\section{Analytical derivation of Eq.~\eqref{eq:threshold}}\label{app:threshold}
We compute the condition for which the payoff for choosing action $+1$ is greater than the payoff for choosing action $-1$. Using~\eqref{eq:payoff_spec1} and~\eqref{eq:payoff_spec2}, we observe that the inequality $\pi_i(+1|y_i,x_{-i})\geq\pi_i(-1|y_i,x_{-i})$ holds if and only if 
\begin{equation}\begin{array}{ll}
    \displaystyle\frac{1}{2}\lambda_i y_i &\displaystyle + \frac{1 - \lambda_i}{2d_i}(1+\alpha)\sum_{j=1}^n a_{ij} (1+x_j)  \\
    & \displaystyle\geq -\frac{1}{2}\lambda_i y_i + \frac{1 - \lambda_i}{2d_i}\sum_{j=1}^n a_{ij} (1-x_j) 
\end{array}\end{equation}
which, after rearranging and recalling that $d_i = \sum_{j=1} a_{ij}$, yields
\begin{equation}
    \frac{1 - \lambda_i}{2d_i}(2+\alpha)\sum_{j=1}^n a_{ij} x_j \geq -\left( \frac{1}{2}(1 -\lambda_i) \alpha + \lambda_i y_i\right).
\end{equation}
The inequality in~\eqref{eq:threshold} can then be recovered 
from the above by further rearranging and simplifying.

\section{Proof of Theorem~\ref{theorem}}\label{app:proof}

 Consider a generic node $i$. According to~\eqref{eq:best_response}, a necessary condition for node $i$ to change action to $+1$ is that $\pi_i(+1; y_i, x_{-i})\geq\pi_i(-1; y_i, x_{-i})$. Using their explicit expression in~\eqref{eq:payoff_spec1} and~\eqref{eq:payoff_spec2}, we bound
\begin{align}
&\pi_i(+1; y_i, x_{-i})\leq\displaystyle \frac12\lambda+\frac{(1-\lambda)(1+\alpha)}{2d_i}\sum_{j=1}^n a_{ij} (1+x_j),\\
&\pi_i(-1; y_i, x_{-i})\geq\displaystyle -\frac12\lambda+\frac{1-\lambda}{2d_i}\sum_{j=1}^n a_{ij} (1-x_j).
\end{align}
In order to start the diffusion, one individual has to adopt $+1$ when all the others (except for the stubborn innovator) take $-1$. We study separately the case in which the first adopter of action $+1$ is a neighbor of the innovator or not. If $i:(i,s)\notin \mathcal E_A$, a necessary (but not sufficient) condition for node $i$ to be the first adopter is derived from the inequalities above as
\begin{equation}
    \frac12\lambda\geq-\frac12+{1-\lambda}\implies \lambda\geq\frac12.
\end{equation}
For a generic individual $i:(i,s)\in\mathcal E_A$, the bounds above yield the following necessary condition for $i$ to be the first adopter of action $+1$:
\begin{equation}
\frac12\lambda+(1-\lambda)(1+\alpha)\frac{1}{d_i}\geq-\frac12\lambda+(1-\lambda)\frac{d_i-1}{d_i},\end{equation}
yielding
\begin{equation}
\lambda(2d_i-2-\alpha)\geq d_i-2-\alpha.\end{equation}
If $d_i> 2+\alpha$, then
\begin{equation}\lambda\geq\frac12-\frac{2+\alpha}{4d_i-4-2\alpha}.
\end{equation}
If $d_i\leq 2+\alpha$, then the necessary condition above is always verified. We observe that the necessary condition for a neighbor of the innovator is always less restrictive than the one for the other individuals, independent of the evolutionary advantage $\alpha$ and of the degree $d_i$. The necessary condition is obtained by minimizing over all the neighbors of the innovator.

\section{Network models and their implementation}\label{app:network}

In the numerical simulations of this paper, we use different network topologies generated according to four different algorithms to obtain a network with $n$ nodes and average degree $d$. Details on the properties of the generated networks can be found in the book by Newman~\cite{newman2010networks_book}, while more details on the specific implementation of these algorithms in this paper are reported in the following.

\begin{description}
\item[Regular random (RR)] the network is generated using a configuration model, that is, each node is given $d$ half-links. A pair of half-links is selected uniformly at random and, if the pair consist of nodes that are not already connected through an edge, then the two half-links are removed and an edge between the two nodes is added. The procedure is repeated until all the half-links are removed.
\item[Erd\H{o}s-R\'enyi (ER)] the network is selected uniformly at random from the ensemble of graphs with $n$ nodes and $dn/2$ edges. This is implemented by selecting a pair of nodes uniformly at random and, if they are not already connected by an edge, adding the edge between them to the edge set. This procedure is repeated until $dn/2$ edges are added to the edge set.
\item[Watts-Strogatz (WS)] the network is generated as follows. First, a regular ring lattice where each node is connected to the $d$ nearest neighbors is constructed. Then, each edge is randomly rewired with probability equal to $p$, independently of the other edges. Edge rewiring is performed by randomly chose one of the two nodes connected by the edge and substituting it with another node, chosen uniformly at random among the other $n-2$ nodes. In all the implementations of WS graphs in this paper, we fix $p=0.2$.
\item[Barab\'asi-Albert (BA)] the network is generated following the preferential attachment algorithm. First, a complete network with $d+1$ nodes is generated. Then, a new node is introduced in the network and $d/2$ edges are generated to connect the new node to $d/2$ existing nodes. Specifically, the probability that the new node is connected with a node $i$ is proportional to the degree of node $i$. The procedure is repeated until the node set contains all the $n$ nodes.
\end{description}

%

\end{document}